\documentstyle[twoside,psfig]{article}

\catcode`\@=11
\long\def\@makefntext#1{
\protect\noindent \hbox to 3.2pt {\hskip-.9pt  
$^{{\eightrm\@thefnmark}}$\hfil}#1\hfill}		

\def\@makefnmark{\hbox to 0pt{$^{\@thefnmark}$\hss}}	
	
\def\ps@myheadings{\let\@mkboth\@gobbletwo
\def\@oddhead{\hbox{}
\rightmark\hfil\eightrm\thepage}   
\def\@oddfoot{}\def\@evenhead{\eightrm\thepage\hfil
\leftmark\hbox{}}\def\@evenfoot{}
\def\sectionmark##1{}\def\subsectionmark##1{}}

\oddsidemargin=\evensidemargin
\newcounter{sectionc}\newcounter{subsectionc}\newcounter{subsubsectionc}
\renewcommand{\section}[1] {\vspace{12pt}\addtocounter{sectionc}{1} 
\setcounter{subsectionc}{0}\setcounter{subsubsectionc}{0}\noindent 
	{\tenbf\thesectionc. #1}\par\vspace{5pt}}
\renewcommand{\subsection}[1] {\vspace{12pt}\addtocounter{subsectionc}{1} 
	\setcounter{subsubsectionc}{0}\noindent 
	{\bf\thesectionc.\thesubsectionc. {\kern1pt \bfit #1}}\par\vspace{5pt}}
\renewcommand{\subsubsection}[1] {\vspace{12pt}\addtocounter{subsubsectionc}{1}
	\noindent{\tenrm\thesectionc.\thesubsectionc.\thesubsubsectionc.
	{\kern1pt \tenit #1}}\par\vspace{5pt}}
\newcommand{\nonumsection}[1] {\vspace{12pt}\noindent{\tenbf #1}
	\par\vspace{5pt}}

\newcounter{appendixc}
\newcounter{subappendixc}[appendixc]
\newcounter{subsubappendixc}[subappendixc]
\renewcommand{\thesubappendixc}{\Alph{appendixc}.\arabic{subappendixc}}
\renewcommand{\thesubsubappendixc}
	{\Alph{appendixc}.\arabic{subappendixc}.\arabic{subsubappendixc}}

\renewcommand{\appendix}[1] {\vspace{12pt}
        \refstepcounter{appendixc}
        \setcounter{figure}{0}
        \setcounter{table}{0}
        \setcounter{lemma}{0}
        \setcounter{theorem}{0}
        \setcounter{corollary}{0}
        \setcounter{definition}{0}
        \setcounter{equation}{0}
        \renewcommand{\thefigure}{\Alph{appendixc}.\arabic{figure}}
        \renewcommand{\thetable}{\Alph{appendixc}.\arabic{table}}
        \renewcommand{\theappendixc}{\Alph{appendixc}}
        \renewcommand{\thelemma}{\Alph{appendixc}.\arabic{lemma}}
        \renewcommand{\thetheorem}{\Alph{appendixc}.\arabic{theorem}}
        \renewcommand{\thedefinition}{\Alph{appendixc}.\arabic{definition}}
        \renewcommand{\thecorollary}{\Alph{appendixc}.\arabic{corollary}}
        \renewcommand{\theequation}{\Alph{appendixc}.\arabic{equation}}
        \noindent{\tenbf Appendix \theappendixc #1}\par\vspace{5pt}}
\newcommand{\subappendix}[1] {\vspace{12pt}
        \refstepcounter{subappendixc}
        \noindent{\bf Appendix \thesubappendixc. {\kern1pt \bfit #1}}
	\par\vspace{5pt}}
\newcommand{\subsubappendix}[1] {\vspace{12pt}
        \refstepcounter{subsubappendixc}
        \noindent{\rm Appendix \thesubsubappendixc. {\kern1pt \tenit #1}}
	\par\vspace{5pt}}

\newcommand{\textlineskip}{\baselineskip=13pt}
\newcommand{\smalllineskip}{\baselineskip=10pt}

\def\eightcirc{
\begin{picture}(0,0)
\put(4.4,1.8){\circle{6.5}}
\end{picture}}
\def\eightcopyright{\eightcirc\kern2.7pt\hbox{\eightrm c}}

\def\abstracts#1#2#3{{
	\centering{\begin{minipage}{4.5in}\footnotesize\baselineskip=10pt
	\parindent=0pt #1\par 
	\parindent=15pt #2\par
	\parindent=15pt #3
	\end{minipage}}\par}} 

\renewenvironment{thebibliography}[1]
	{\frenchspacing
	 \ninerm\baselineskip=11pt
	 \begin{list}{\arabic{enumi}.}
	{\usecounter{enumi}\setlength{\parsep}{0pt}
	 \setlength{\leftmargin 12.7pt}{\rightmargin 0pt} 
	 \setlength{\itemsep}{0pt} \settowidth
	{\labelwidth}{#1.}\sloppy}}{\end{list}}

\newcounter{itemlistc}
\newcounter{romanlistc}
\newcounter{alphlistc}
\newcounter{arabiclistc}

\newcommand{\fcaption}[1]{
        \refstepcounter{figure}
        \setbox\@tempboxa = \hbox{\footnotesize Fig.~\thefigure. #1}
        \ifdim \wd\@tempboxa > 5in
           {\begin{center}
        \parbox{5in}{\footnotesize\smalllineskip Fig.~\thefigure. #1}
            \end{center}}
        \else
             {\begin{center}
             {\footnotesize Fig.~\thefigure. #1}
              \end{center}}
        \fi}

\newcommand{\tcaption}[1]{
        \refstepcounter{table}
        \setbox\@tempboxa = \hbox{\footnotesize Table~\thetable. #1}
        \ifdim \wd\@tempboxa > 5in
           {\begin{center}
        \parbox{5in}{\footnotesize\smalllineskip Table~\thetable. #1}
            \end{center}}
        \else
             {\begin{center}
             {\footnotesize Table~\thetable. #1}
              \end{center}}
        \fi}

\def\@citex[#1]#2{\if@filesw\immediate\write\@auxout
	{\string\citation{#2}}\fi
\def\@citea{}\@cite{\@for\@citeb:=#2\do
	{\@citea\def\@citea{,}\@ifundefined
	{b@\@citeb}{{\bf ?}\@warning
	{Citation `\@citeb' on page \thepage \space undefined}}
	{\csname b@\@citeb\endcsname}}}{#1}}

\newif\if@cghi
\def\cite{\@cghitrue\@ifnextchar [{\@tempswatrue
	\@citex}{\@tempswafalse\@citex[]}}
\def\citelow{\@cghifalse\@ifnextchar [{\@tempswatrue
	\@citex}{\@tempswafalse\@citex[]}}
\def\@cite#1#2{{$\null^{#1}$\if@tempswa\typeout
	{IJCGA warning: optional citation argument 
	ignored: `#2'} \fi}}

\def\pmb#1{\setbox0=\hbox{#1}
	\kern-.025em\copy0\kern-\wd0
	\kern.05em\copy0\kern-\wd0
	\kern-.025em\raise.0433em\box0}

\def\fnt#1#2{\footnotetext{\kern-.3em
	{$^{\mbox{\scriptsize #1}}$}{#2}}}

\def\@makefnmark{\hbox to 0pt{$^{\@thefnmark}$\hss}}	
	
\def\ps@myheadings{%
    \let\@oddfoot\@empty\let\@evenfoot\@empty
    \def\@evenhead{\slshape\leftmark\hfil}
    \def\@oddhead{\hfil{\slshape\rightmark}}
    \let\@mkboth\@gobbletwo
    \let\sectionmark\@gobble
    \let\subsectionmark\@gobble
    }
\font\tenrm=cmr10
\font\tenit=cmti10 
\font\tenbf=cmbx10
\font\bfit=cmbxti10 at 10pt
\font\ninerm=cmr9

\font\eightrm=cmr8

\def\qed{\hbox{${\vcenter{\vbox{			
   \hrule height 0.4pt\hbox{\vrule width 0.4pt height 6pt
   \kern5pt\vrule width 0.4pt}\hrule height 0.4pt}}}$}}

\begin{document}
\def \beq {\begin{equation}}
\def \eeq {\end{equation}}
\def \bes {\begin{eqnarray}}
\def \ees {\end{eqnarray}}
\def\ni{\noindent}
\def\nn{\nonumber}
\def\rv{\mbox{\boldmath$r$}}
\def\kv{\mbox{\boldmath$k$}}
\def\drv{{d\mbox{\boldmath$r$}}}
\def\mum {$\,\mu\mbox{m}$}
\def\vkp{{\mbox{\boldmath{$k$}}}_{\bot}}
\def\e{\varepsilon}
\def\kp{k_{\bot}}
\setlength{\textheight}{7.7truein}  

\thispagestyle{empty}

\normalsize\textlineskip

\setcounter{page}{1}

\vspace*{0.88truein}
\centerline{\bf WHAT IS THE TEMPERATURE DEPENDENCE}
\vspace*{0.035truein}
\centerline{\bf  OF THE CASIMIR FORCE BETWEEN REAL METALS?
}
\vspace*{0.37truein}
\centerline{\footnotesize G. L. KLIMCHITSKAYA\footnote{On leave
from North-West Polytechnical University, St.Petersburg, Russia}}
\baselineskip=12pt
\centerline{\footnotesize\it Department of Physics, 
Federal University of Para\'{\i}ba}
\baselineskip=10pt
\centerline{\footnotesize\it Caixa Postal 5008, CEP 58059--970,
Jo\~{a}o Pessoa, Pb-Brazil\footnote{E-mail: galina@fisica.ufpb.br}}
\vspace*{0.225truein}

\vspace*{0.21truein}
\abstracts{The situation with the temperature corrections to
the Casimir force between real metals of finite conductivity
is reported. It is shown that the plasma dielectric function
is well adapted to the Lifshitz formula and leads to
reasonable results for real conductors. The Drude dielectric
function which describes media with dissipation is found
not to belong to the application range of the Lifshitz
formula at nonzero temperature. For Drude metals the
special modification of the zero-frequency term of this
formula is suggested. The contradictory results on the
subject in recent literature are analysed and explained.
}{}{}


\vspace*{1pt}\textlineskip	
\section{Introduction}	
\vspace*{-0.5pt}
\noindent
Currently much attention is given to the Casimir effect\cite{1}
and its topical applications in both fundamental 
physics and nanotechnology.
This effect implies that there is some force acting between two
uncharged bodies closely spaced in vacuum which is caused by 
zero-point oscillations of electromagnetic field. New precision
measurements of the Casimir force have spurred the development
of more exact theoretical methods taking into account such
relevant factors as finite conductivity of the boundary material,
nonzero temperature and surface roughness (for a review of modern
experimental and theoretical developments in the Casimir effect
see Ref.~2).

This paper is devoted to the investigation of the temperature
dependence of the Casimir force acting between real metals, and
one might believe that the question in the title has an apparent
answer. There is the famous Lifshitz formula\cite{3,4}
in the form of a frequency sum for the Casimir force between
two dielectric semispaces separated by a gap at nonzero
temperature. Although there was a problem in application of
this formula to planes made of ideal metals (the zero-frequency
term becomes indefinite for the infinitely large dielectric
permittivity), Schwinger, DeRaad and Milton\cite{5} 
have long demonstrated how to proceed in the case of ideal
metal. One must take the limit of infinite dielectric permittivity
before putting frequency equal to zero. In doing so, the
result for a perfect conductor is obtained from the Lifshitz
theory. It is in agreement with the Casimir force calculated within
the limits of quantum field theory with Dirichlet boundary
conditions in terms of the free energy density of vacuum\cite{6,7}.
One would expect that in order to calculate the Casimir force
between real metals it is suffice to substitute some model
dependence of a metallic dielectric permittivity on frequency
into the Lifshitz formula and make all summations and
integrations correctly.

This approach, however, faces with serious problems. Starting
in 2000, several theoretical groups undertook a number of
studies of the Casimir force at nonzero temperature between
real metals on the base of the Lifshitz formula. In Refs.~8,\,9
the Drude model was used to describe the dependence of the
dielectric permittivity on frequency whereas in Ref.~10 the
dielectric permittivity was described by the plasma model.
In Refs.~11--13 both the plasma and Drude models were used 
on the base of the Lifshitz formula. The zero-frequency term
of this formula was modified in Refs.~11--13 in the same way
as for a perfect conductor\cite{5}. The results of Refs.~8,\,9
and 11--13 run into obstacles and are found to be in 
contradiction with the fundamental physical principles and
experiment (see below). In Refs.~14--16 the Casimir force was
also computed by the use of both plasma and Drude models.
The results obtained in framework of the plasma model
were found to be in agreement with those of Ref.~10.
As for the Drude model, the new prescription for the
zero-frequency term of the Lifshitz formula was 
proposed\cite{15} which makes it possible to avoid the
nonphysical results of Refs.~8,\,9,\,11--13.

In this paper we report the present situation in the problem
of calculation of the Casimir force between real metals
at nonzero temperature. It is shown that in Refs.~8,\,9
the Lifshitz formula was applied outside of its application
range, whereas in Refs.~11--13 the unjustified prescription
for the zero-frequency term of this formula was used.
In Sec.~2 the brief formulation of the Lifshitz formula is
presented. Sec.~3 contains discussion of the application range 
of the Lifshitz formula starting from the modern derivation
in the framework of Quantum Field Theory at nonzero temperature
in Matsubara formulation. In Sec.~4 the results obtained in the
framework of the plasma model are presented. Sec.~5 is devoted
to the case of Drude model. In Sec.~6 the reader will find
conclusions.

\section{The Lifshitz formula}	
\vspace*{-0.5pt}
\noindent
The original Lifshitz formula describing the Casimir and van der
Waals force acting between two semispaces  with
a dielectric permittivity $\varepsilon(\omega)$ at a temperature $T$
separated by a gap of width $a$ 
can be represented in the form of the sum over discrete frequencies
\bes
&&
F_{ss}(a)=-\frac{k_BT}{2\pi}
\sum\limits_{l=-\infty}^{\infty}
\int_{0}^{\infty}k_{\bot}\,dk_{\bot}\,q_l
\left\{\left[r_1^{-2}(\xi_l,k_{\bot})e^{2aq_l}-1\right]^{-1}\right.
\nn \\
&&\phantom{aaaaaaaaaaaaaaaaaaaaa}
+\left.\left[r_2^{-2}(\xi_l,k_{\bot})e^{2aq_l}-1\right]^{-1}\right\},
\label{1}
\ees
\ni
where $r_{1,2}$ are the reflection coefficients with parallel
(perpendicular) polarization, respectively, given by
\beq
r_1^{-2}(\xi_l,k_{\bot})=\left[
\frac{\varepsilon(i\xi_l)q_l+k_l}{\varepsilon(i\xi_l)q_l-k_l}\right]^2,
\qquad
r_2^{-2}(\xi_l,k_{\bot})=\left(
\frac{q_l+k_l}{q_l-k_l}\right)^2.
\label{2}
\eeq
\ni
Here ${\kv}_{\bot}$ is the momentum component lying in the boundary planes,
$k_{\bot}=|{\kv}_{\bot}|$, $\omega=i\xi$, and the following notations
are used
\beq
q_l=\sqrt{\frac{\xi_l^2}{c^2}+k_{\bot}^2}, \quad
k_l=\sqrt{\varepsilon(i\xi_l)\frac{\xi_l^2}{c^2}+k_{\bot}^2},
\quad
\xi_l=\frac{2\pi l}{\beta},
\quad 
\beta\equiv\frac{\hbar}{k_B T},
\label{3}
\eeq
\ni
$k_B$ being the Boltzmann constant.

In the limit of zero temperature the Lifshitz formula can be written
in terms of integrals
\bes
&&
F_{ss}(a)=-\frac{\hbar}{2\pi^2}
\int_{0}^{\infty}d\xi
\int_{0}^{\infty}k_{\bot}\,dk_{\bot}\,q
\left\{\left[r_1^{-2}(\xi,k_{\bot})e^{2aq}-1\right]^{-1}\right.
\nn \\
&&\phantom{aaaaaaaaaaaaaaaaaaaaa}
+\left.\left[r_2^{-2}(\xi,k_{\bot})e^{2aq}-1\right]^{-1}\right\}.
\label{4}
\ees

It is significant that both Eqs.~(\ref{1}) and (\ref{4}) were
originally derived for the case of nondissipative dielectric
media. As was shown later\cite{17,18} by the consideration
of an auxiliary electrodynamic problem, at zero temperature
the Lifshitz formula given by Eq.~(\ref{4}) preserves its validity
even for media with dissipation (see Ref.~19 for details). Below we will
make sure that this important conclusion can not be extended for
the case of nonzero temperature. As a result, the Lifshitz formula
(\ref{1}) at $T>0$ can not be applied in the case of dissipative
media without the appropriate modification of its zero-frequency
term (see also Ref.~20)..

\section{Application range of the Lifshitz formula}	
\vspace*{-0.5pt}
\noindent
Recently the new derivation of the Lifshitz formula was 
performed\cite{2} in the framework of Quantum Field Theory
at nonzero temperature in Matsubara formulation.
In this approach one considers Euclidean field theory with
the electromagnetic field periodic in the Euclidean time
variable within the interval $\beta$. For two semispaces
the calculation of the free energy density is reduced to
the solution of a one-dimensional scattering problem.
The result is\cite{2}
\beq
E_{ss}(a)=-\frac{\hbar}{2\beta}
\sum\limits_{l}\int
\frac{d\vkp}{(2\pi)^2}\left[
\ln s_{11}^{||}\left(i\xi_l,\vkp
\right)+
\ln s_{11}^{\bot}\left(i\xi_l,\vkp
\right)\right],
\label{5}
\eeq
\ni
where $s_{11}^{||}\left(i\xi_l,\vkp\right)$ and
 $s_{11}^{\bot }\left(i\xi_l,\vkp\right)$ 
are the scattering coefficients for parallel and
perpendicular polarizations, respectively.
 The solution of the
scattering problem reads 
\bes
&&
s_{11}^{||}=
\frac{4\e(i\xi_l)k_lq_le^{k_la}}{\left[\e(i\xi_l)q_l+
k_l\right]^2e^{q_la}-\left[\e(i\xi_l)q_l-k_l\right]^2e^{-q_la}},
\nn \\
&&
s_{11}^{\bot}=
\frac{4k_lq_le^{k_la}}{\left(q_l+
k_l\right)^2e^{q_la}-\left(q_l-k_l\right)^2e^{-q_la}},
\label{6}
\ees
\ni
which is valid only if $q_l\neq k_l$ (see Eq.~(\ref{3}) for
notations). If, however, $q_l=k_l$ the element of the scattering
matrix $s_{11}^{\bot}$ proves to be arbitrary and the
non-diagonal element $s_{12}^{\bot}=0$. What this means is that
the scattering problem may not have any definite solution
in the case when $q_l=k_l$. As discussed below, this fact is of 
crucial importance for the determination of the application range
of the Lifshitz formula.

If $q_l\neq k_l$ and the scattering problem has definite solution,
one can substitute Eq.~(\ref{6}) into Eq.~(\ref{5}) in order
to get the free energy density. After performing renormalization,
which is equivalent to the omitting of the free energy in the
case of infinitely remote plates, one obtains\cite{2}
\bes
&&
E_{ss}(a)=\frac{k_B T}{4\pi}
\sum\limits_{l}
\int_{0}^{\infty}
\kp d\kp
\left\{
\ln\left[1- r_{1}^2\left(\xi_l,\kp\right)
e^{-2aq_l}\right]\right.
\nn \\
&&\phantom{aaaaaaaaaaaaaaaaaa}
\left.+
\ln\left[1- r_{2}^2\left(\xi_l,\kp\right)
e^{-2aq_l}\right]
\right\}.
\label{7}
\ees
\ni
It is easily seen that the Lifshitz formula (\ref{1})
is obtained from Eq.~(\ref{7}) as the minus derivative
of (\ref{7}) with respect to $a$. Note that according
to the Proximity Force Theorem\cite{20} the force acting
between a semispace (plate) and a sphere (spherical lens)
can be calculated approximately as
\beq
F_{sl}(a)=2\pi R E_{ss}(a).
\label{8}
\eeq

To discuss the application range of Eqs.~(\ref{1}), (\ref{7})
let us start from the plasma model representation for the
dielectric permittivity
\beq
\varepsilon(\omega)=1-\frac{\omega_p^2}{\omega^2},
\quad
\varepsilon(i\xi)=1+\frac{\omega_p^2}{\xi^2},
\label{9}
\eeq
\ni
where $\omega_p$ is the plasma frequency.
If to substitute Eq.~(\ref{9}) into  Eq.~(\ref{3})
one makes sure that $q_l\neq k_l$ including the
zero-frequency case where
\beq
q_0=k_{\bot},\qquad
k_0=\sqrt{\frac{\omega_p^2}{c^2}+k_{\bot}^2}.
\label{10}
\eeq
\ni
The respective reflection coefficients at zero frequency are
\beq
r_1^2(0,k_{\bot})=1, \quad
r_2^2(0,k_{\bot})=\left(\frac{k_{\bot}-\sqrt{k_{\bot}^2+
\frac{\omega_p^2}{c^2}}}{k_{\bot}+\sqrt{k_{\bot}^2+
\frac{\omega_p^2}{c^2}}}\right)^2.
\label{11}
\eeq
\ni
Note that the first of them is equal to the reflection
coefficient of real photons at zero frequency 
in the case of metals and the second one is smaller than unity.
It is seen that the plasma model belongs to the application
range of the Lifshitz formula and no difficulties arise.
In the case of $\omega_p\to\infty$ one obtains
\beq
\lim\limits_{\omega_p\to\infty}
r_1^2(\xi_l,k_{\bot})=
\lim\limits_{\omega_p\to\infty}
r_2^2(\xi_l,k_{\bot})=1,
\label{12}
\eeq
\ni
i.e. the limit of ideal metal as it should be expected
from physical considerations. In the next section some 
calculational results obtained in the framework of the 
plasma model are presented.

It is appropriate at this point to consider the cases
when $q_l=k_l$ and the scattering problem may not have 
definite solution. Unexpectedly, this is the case for
the nondissipative dielectric media described by
\beq
\varepsilon (\omega)=
\varepsilon (i\xi)=\varepsilon_0=\mbox{const},
\label{13}
\eeq
\ni
where at zero frequency $q_0=k_0=k_{\bot}$. In spite of
this, some additional considerations help to fix the
solution of the scattering problem for dielectrics.
By the use of the unitarity condition which is valid
for nondissipative media one arrives at
$|s_{11}^{\bot}(0,k_{\bot})|=1$, 
and after the application of the dispersion 
relation\cite{2} the phase of the scattering matrix element 
can be also fixed with the result 
$s_{11}^{\bot}(0,k_{\bot})=1$.
It is significant that exactly the same result is
obtainable from the general solution (\ref{6}) of the
scattering problem in a formal limit
$\xi\to 0$. 
Thus, dielectrics turned out to belong to the application
range of the Lifshitz formula along with metals
described by the plasma model.

The values of reflection coefficients for dielectrics are
found from Eqs.~(\ref{2}), (\ref{13})
\beq
r_1^2(0,k_{\perp})=\left(\frac{\varepsilon_0
-1}{\varepsilon_0+1}\right)^2,
\quad
r_2^2(0,k_{\perp})=0.
\label{16}
\eeq
\ni
It is notable that they do not coincide with the
values of reflection coefficients for real photons
given by
\beq
r_{1(R)}^2=r_{2(R)}^2\equiv r_{(R)}^2
=
\left(\frac{\sqrt{\varepsilon_0}
-1}{\sqrt{\varepsilon_0}+1}\right)^2.
\label{17}
\eeq
\ni 
As is seen from (\ref{16}), (\ref{17}),
$r_1>r_{(R)}$ and $r_2<r_{(R)}$.
The computational results for dielectrics are given below
in Sec.~5.

The other possibility when $q_l=k_l$ and the scattering
problem may not have a solution takes place for metals
described by the Drude dielectric function
\beq
\varepsilon(\omega)=
1-\frac{\omega_p^2}{\omega(\omega+i\gamma)},
\quad
\varepsilon(i\xi)=1+\frac{\omega_p^2}{\xi(\xi+\gamma)},
\label{18}
\eeq
\ni
where $\gamma$ is the relaxation frequency. In fact here at
zero frequency 
$q_0=k_0=k_{\bot}$, i.e. the same as for dielectrics.

In this case, however, the unitarity condition is absent
because of nonzero dissipation. As a result there is no way
to solve the scattering problem at zero frequency and the 
matrix element 
$s_{11}^{\bot}(0,k_{\bot})$ and thereby the reflection
coefficient
$r_2(0,k_{\bot})$ remain undetermined.
The conclusion is that real metals described by the
Drude dielectric function are outside of the application
range of the Lifshitz formula as given by Eq.~(\ref{1}).

For this reason it is of dubious value to substitute the
reflection coefficients at zero frequency
\beq
r_1^2(0,k_{\bot})=1, \quad
r_2^2(0,k_{\bot})=0,
\label{19}
\eeq
\ni
following from Eqs.~(\ref{2}), (\ref{18}) in the case of the
Drude model, into the Lifshitz formula, as was done in
Refs.~8,\,9. The temperature corrections to the Casimir and
van der Waals force computed in such a way may be and really 
are in contradiction with both the most fundamental
physical principles and also with experiment (see Sec.~5).

To compute the Casimir and van der Waals force at nonzero
temperature for real metal as described by the Drude model,
some modification of the zero-frequency term of the Lifshitz 
formula is required. Such modification was attempted in
Refs.~11--13 where the following values of the reflection
coefficients were postulated for both plasma and Drude models
\beq
r_1^2(0,k_{\bot})=r_2^2(0,k_{\bot})=1
\label{20}
\eeq
\ni
as given by the perfect conductors. This prescription is
also shown to be in contradiction with both the general 
principles of thermodynamics and with experiment (see
Secs.~4,\,5). The more adequate prescription applicable
in the case of the Drude model and leading to no changes in
the case of the plasma model was proposed in Ref.~15 and 
is discussed in Sec.~5.

\section{Temperature Casimir force in the plasma model}	
\vspace*{-0.5pt}
\noindent
As was discussed in the preceding section, for real metals,
as described by the plasma model, the Lifshitz formula is
well defined and does not need any modification.
The results for the force per unit area and free energy
density between two semispaces are given by Eqs.~(\ref{1}),
(\ref{7}), where the reflection coefficients are defined
in (\ref{2}) and dielectric permittivity --- in (\ref{9}).
The force between a semispace and a sphere can be obtained
from Eq.~(\ref{8}). The computations in the framework
of the plasma model were performed in Refs.~10,\,14,\,16.
In the case of low temperatures ($T\ll T_{eff}$ where
$k_B T_{eff}=\hbar c/(2a)$), or, equivalently, small
separations the result can be obtained perturbatively\cite{14} 
\bes
&&
E_{ss}(a)\approx
-\frac{\pi^2\hbar c}{720a^3}\left\{1+
\frac{45\zeta(3)}{\pi^3}\left(\frac{T}{T_{eff}}\right)^3
-\left(\frac{T}{T_{eff}}\right)^4\right.
\nn \\
&&\phantom{aa}
\left.
-4\frac{\delta_0}{a}\left[1-
\frac{45\zeta(3)}{2\pi^3}\left(\frac{T}{T_{eff}}\right)^3
+\left(\frac{T}{T_{eff}}\right)^4\right]\right\},
\label{21}
\ees
\ni
where $\delta_0=c/\omega_p$ is the effective penetration
depth of electromagnetic zero-point oscillations into a metal.

For the force acting between two semispaces the result 
is\cite{14}
\beq
F_{ss}(a)\approx 
-\frac{\pi^2\hbar c}{240a^4}\left\{1+
\frac{1}{3}\left(\frac{T}{T_{eff}}\right)^4
-\frac{16}{3}\frac{\delta_0}{a}
\left[1-\frac{45\zeta(3)}{8\pi^3}
\left(\frac{T}{T_{eff}}\right)^3\right]\right\}.
\label{22} 
\eeq
\ni
It is notable that at low temperatures (small separations)
the temperature correction in Eqs.~(\ref{21}), (\ref{22})
depends only on the third and fourth powers in $T/T_{eff}$.
This is in contradiction with Refs.~11--13 where for real 
metals, as described by the plasma model, the linear in
temperature correction arises at low temperatures.
Evidently, linear temperature correction leads to nonzero 
value of entropy at zero temperature. This value depends on
the parameters of 
the system like the plasma frequency and space separation and, 
thereby, is in contradiction with the third law of
thermodynamics (Nernst heat theorem). Because of this,
prescription (\ref{20}) used in Refs.~11--13
is unacceptable in the case of metals
described by the plasma model.

In the case of high temperatures (large separations) the 
results for the free energy density and force obtained on the
base of the Lifshitz formula and plasma model are\cite{14}
\beq
E_{ss}(a)=-\frac{k_BT}{8\pi a^2}\zeta(3)
\left(1-2\frac{\delta_0}{a}\right),
\quad
F_{ss}(a)=-\frac{k_BT}{4\pi a^3}\zeta(3)
\left(1-3\frac{\delta_0}{a}\right),
\label{23}
\eeq
\ni
where $\zeta(z)$ is Riemann zeta-function.

In contrast to Eq.~(\ref{23}), at large separations the results
of Refs.~11--13 contain no finite conductivity corrections,
i.e. are given by the first contributions in the right-hand
sides of (\ref{23}). By way of example, at $T=300\,$K
the finite conductivity corrections computed on the basis of
Refs.~11--13 become zero at separations $a\geq 5\,\mu$m 
regardless of the quality of a metal under consideration which
is a nonphysical property.

\section{Temperature Casimir force in the Drude model}	
\vspace*{-0.5pt}
\noindent
As was shown in Sec.~3, the Drude dielectric function, which
describes media with dissipation, is outside of the application
range of the Lifshitz formula (\ref{1}). To substitute the Drude 
dielectric function into the Lifshitz formula the special prescription
must be adopted redefining its zero-frequency term. This should be
done by following the general physical requirements (like the laws
of thermodynamics) and preserving the solid results obtained earlier in
the limiting cases of ideal metal and plasma model. Prescription of
this kind was suggested in Ref.~15. It can be formulated as follows.

The zero-frequency term of the Lifshitz formula (\ref{5})
for the free energy density can be represented as
\bes
&&
E_{ss}^{(l=0)}(a)=\frac{k_B T}{16\pi a^2}
\int_0^{\infty}y\,dy\left\{
\vphantom{\int_0^{\infty}}
\ln\left[1-r_1^2(0,y)e^{-y}\right]\right.
\nn \\
&&
\phantom{aaa}
+\left.
\ln\left[1-r_2^2(y,y)e^{-y}\right]-
\int_0^{y} dx\frac{\partial}{\partial x}
\ln\left[1-r_2^2(x,y)e^{-y}\right]\right\}.
\label{24}
\ees
\ni
To obtain Eq.~(\ref{24}) the continuous frequency $\xi$ was
introduced in (\ref{3}) and then the dimensionless variables
were considered
\beq
x=2a\frac{\xi}{a}, \quad
y=2a\sqrt{\frac{\xi^2}{c^2}+k_{\bot}^2}.
\label{25}
\eeq
\ni
The derivative with respect to $x$ in (\ref{24}) is identically
equal to zero in the case of the plasma model.
In the case of the Drude model, however, this derivative
was proved to be a discontinuous function of the relaxation
parameter $\gamma$.\cite{15}
As a result, the presence of this term in Eq.~(\ref{24}) renders
it unsuitable for real metals described by the Drude model.

To improve the situation one may use the fact that
\beq
\frac{\partial\ln\left[1-r_2^2(x,y)e^{-y}\right]}{\partial x}
\sim\frac{1}{\sqrt{\varepsilon -1}}\to 0
{\ \mbox{with}\ }\varepsilon\to\infty.
\label{26}
\eeq
\ni
Due to this the discontinuous term can be deleted by the
prescription of Ref.~5. As a result, the modified zero-frequency
term of the Lifshitz formula is given by
\bes
&&
E_{ss}^{(l=0)}(a)=\frac{k_B T}{16\pi a^2}
\int_0^{\infty}y\,dy\left\{
\ln\left[1-r_1^2(0,y)e^{-y}\right]\right.
\nn \\
&&
\phantom{aaa}
+\left.
\ln\left[1-r_2^2(y,y)e^{-y}\right]
\right\}
\label{27}
\ees
\ni
with all the other terms the same as in Eq.~(\ref{5}).

\begin{figure}[tb]
\vspace*{-10.cm}
\centerline{\psfig{file=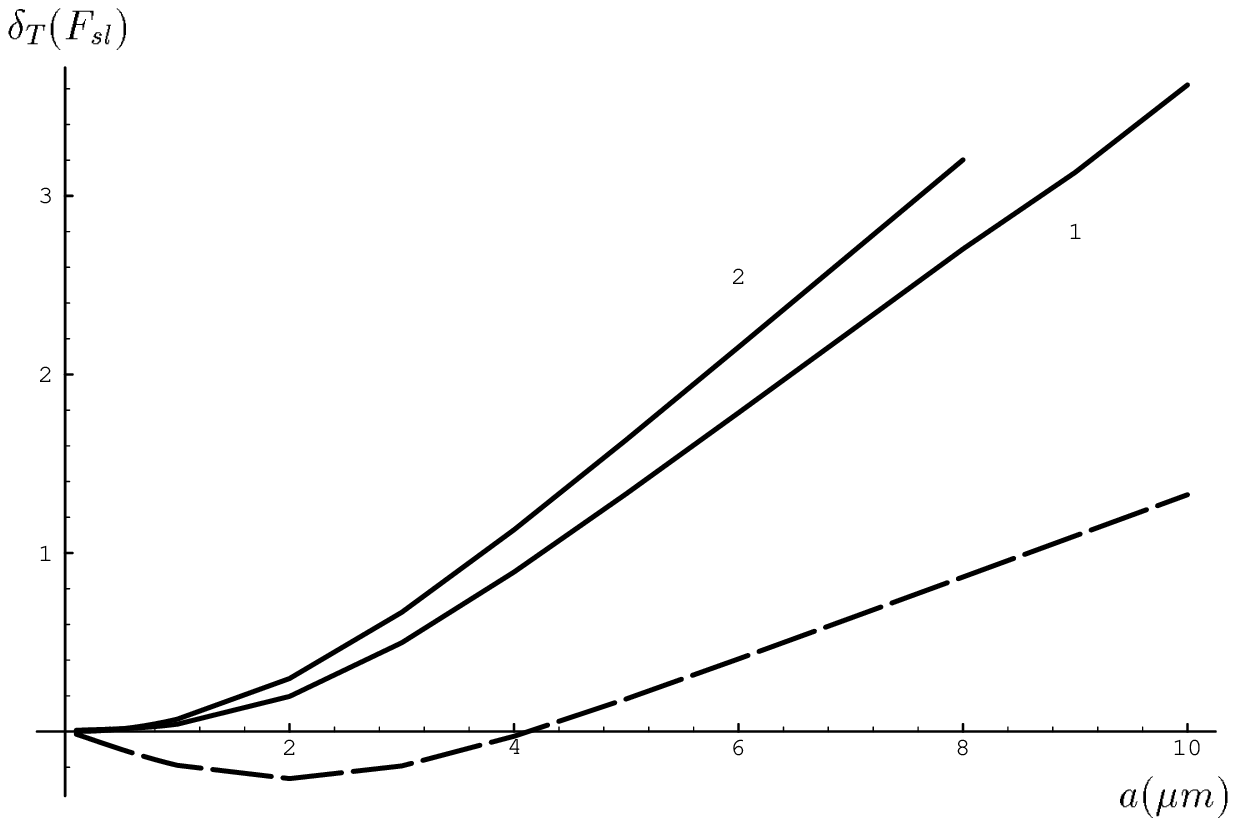}} 
\vspace*{-10.5cm}
\fcaption{\label{energyT}
Relative temperature correction to the Casimir
force between a plate and a lens in dependence of 
separation.
Curve 1 corresponds to Drude model (our computation),
the dashed curve is obtained in Drude model with
$r_2(0,k_{\perp})=0$, and curve 2 is for the dielectric
test bodies.
}
\end{figure}

By way of example, in Fig.~1 the relative temperature correction
to the free energy density between plates is presented 
at $T_0=300\,$K (which is
the same as the relative temperature correction to the force
between a sphere and a plate)
\beq
\delta_T(E_{ss})=\delta_T(F_{sl})=
\frac{E_{ss}(a,T_0)-E_{ss}(a,0)}{E_{ss}(a,0)}.
\label{28}
\eeq
\ni
The results obtained by the Lifshitz formula with a modified
zero-frequency term (\ref{27}) are shown by curve 1
($\omega_p=12.5\,$eV, $\gamma=0.063\,$eV as for $Al$),
curve 2 is calculated for dielectrics with $\varepsilon_0=7$,
the dashed curve is obtained in the framework of Drude model
and unmodified Lifshitz formula\cite{9} (i.e. by the use of
Eq.~(\ref{19})). Both curves 1 and 2 demonstrate
reasonable behavior. As to the dashed curve, it clearly
demonstrates negative temperature correction at small
and moderate separations. At $a=1\,\mu$m this correction
corresponds to about 17\% of the force which is in
contradiction with experiment\cite{21}. Note that the 
negative temperature corrections are unacceptable from
the theoretical point of view because they imply that the entropy
of a system of two plates 
is negative (in this case within a separation range 
$0<a<4\,\mu$m). At high temperatures the result of Refs.~8,\,9
demonstrates only one half of the asymptotic value for a perfect
conductor irrespective of how high the conductivity of real
metal is which is a nonphysical property.

Analogical results are obtained for the force between two plates\cite{15}.
The negative temperature correction to the Casimir force
obtained in Refs.~8,\,9 is also in contradiction with the evident
physical argument that with an increase of temperature the
population of all modes, and thereby force modulus, increase.

The results obtained in Refs.~11--13 on the base of Drude model
with a prescription (\ref{20}) are also nonphysical. Here, once
more, a linear (although positive) temperature correction arises
at small separations. It is in contradiction with both the
Nernst heat theorem and experiment\cite{22}. At large
separations the approach of Refs.~11--13 does not permit to describe
the finite conductivity corrections to the Casimir force between real
metals.

\section{Conclusions}	
\vspace*{-0.5pt}
\noindent
The following conclusions can be formulated from the above
considerations.

1.
The plasma model is well adapted for the description of the Casimir
force between real metals at nonzero temperature. The Lifshitz
formula with the plasma dielectric function is completely consistent
mathematically and is not a subject for any modifications.

2.
The Drude dielectric function describing the media with dissipation is
outside of the application range of the Lifshitz formula at nonzero 
temperature. The zero frequency term of the Lifshitz formula with the
Drude dielectric function remains indeterminate.

3.
We propose the redefinition of the zero-frequency term of the Lifshitz
formula in the presence of dissipation which is in accordance with the
principles of thermodynamics and leads to physically consistent results.

4.
The rigorous derivation of the Lifshitz formula at nonzero temperature
in the case of dissipative media (including the zero-frequency term)
is the important problem to be solved in near future.

\nonumsection{Acknowledgments}
\noindent
The author is grateful to M.~Bordag and V.M.~Mostepanenko for helpful
discussions. She is indebted to the Organizers of the Fifth Workshop on
Quantum Field Theory under the Influence of 
External Conditions (Leipzig University, 2001)
for kind hospitality. The financial support from CNPq is acknowledged.

\nonumsection{References}

\end{document}